\documentclass[pra,aps,superscriptaddress,showpacs,amsmath,amssymb,twocolumn]{revtex4}

\usepackage{graphicx}

\newcommand{\xt}{\ensuremath{\breve{\mathbf{x}}}_{t}}
\newcommand{\xo}{\ensuremath{\breve{\mathbf{x}}}_{0}}
\newcommand{\qt}{\ensuremath{\breve{q}}_{t}}
\newcommand{\pt}{\ensuremath{\breve{p}}_{t}}
\newcommand{\Xt}{\ensuremath{\breve{X}_t}}
\newcommand{\Et}{\ensuremath{\breve{E}_t}}
\newcommand{\xop}{\ensuremath{\mathbf{\hat x}}}
\renewcommand{\a}{\ensuremath{\alpha}}
\newcommand{\Tr}[1][0]{\ensuremath{\mbox{Tr}_{}\left[ {#1} \right]}}

\begin{document}

\title{Effects of measurement back-action in the stabilization of a Bose-Einstein condensate through feedback}
\author{S.D. Wilson}
\email{stuart.wilson@anu.edu.au} \affiliation{Department of Engineering, Australian National University, Canberra, ACT 0200, Australia.}
\author{A.R.R. Carvalho}
\affiliation{Department of Physics, Australian National University, Canberra, ACT 0200, Australia.}
\author{J.J. Hope}
\affiliation{Australian Centre for Quantum-Atom Optics, Department of Physics, Australian National University, Canberra, ACT 0200, Australia.}
\author{M.R. James}
\affiliation{Department of Engineering, Australian National University, Canberra, ACT 0200, Australia.}

\date{\today}

\begin{abstract}

We apply quantum filtering and control to a particle in a harmonic trap under continuous position measurement, and show that a simple static feedback law can be used to cool the system. The final steady state is Gaussian and dependent on the feedback strength and coupling between the system and probe. In the limit of weak coupling this final state becomes the ground state. An earlier model by Haine \emph{et. al.} (PRA, {\bf 69} 2004) without measurement back-action showed dark states: states that did not display error signals, thus remaining unaffected by the control. This paper shows that for a realistic measurement process this is not true, which indicates that a Bose-Einstein condensate may be driven towards the ground state from any arbitrary initial state. 

\end{abstract}

\pacs{03.75.Hh,02.30.Yy,02.50.-r,42.50.Lc}

\maketitle

\section{Introduction}

The observation and control of a Bose-Einstein condensate (BEC) using a real-time feedback loop has become an important scientific objective within the atom optics community.  Such a process may be essential to reduce the noise characteristics of an atom laser that is outcoupled from such a BEC.  There are two classes of noise that require control.  The first is the quantum statistical noise of the BEC mode, which limits the linewidth of an atom laser outcoupled from a single mode BEC \cite{wise_thomsen_01,Johnsson_05}.  The second is the spatial mode of the BEC itself, which may not be stable under varying preparation conditions \cite{corney99}, or continuous pumping \cite{haine04,Johnsson_05}.  The quantum statistical noise of the mode will not be the dominant effect on the properties of the atom laser unless the BEC can be produced in a single spatial mode, so the first consideration of implementing control schemes will be to provide that stable spatial mode.  This paper shows that a straightforward static measurement feedback can address that problem.

In previous work, Haine \textit{et al.} address controlling a BEC system with and without nonlinearities using realistic trap controls, but did not model the back-action of their measurement system \cite{haine04}.  They model the system semiclassically, and provide feedback based on the state of the system, rather than include the effects of coupling the BEC to a measurement device.  In the absence of atomic interactions, this work predicted non-Gaussian {\lq dark states\rq} that did not respond to the control, although they were not in the ground state.  Doherty and Jacobs include this back-action in a model of the measurement of a single atom trapped in a harmonic potential, and they provide an optimal control scheme for driving it towards the ground state \cite{doherty99}.  This result only holds for a Gaussian initial state, so even in the absence of atomic interactions, the dark states of the semiclassical treatment could not be addressed by that analysis.  Position measurement and control schemes for Gaussian quantum systems have also been considered for nanomechanical resonators \cite{Hopkins} and atomic cavity QED \cite{Steck}.

In this paper we show that a single trapped particle in an arbitrary quantum state can always be cooled down to the ground state by the experimentally realistic feedback scheme shown in reference \cite{haine04}, as measurement back-action destabilises the dark states shown in that analysis.

In Sec.~\ref{Model}, we introduce our model of the system and the quantum filter that is used to provide the control.  Sec.~\ref{FEE} produces analytic results for the final expected energy of the system under the influence of feedback.  It does this first for a Gaussian initial state, and then shows that the final state is independent of the statistics of the initial state.  Sec.~\ref{Sim} shows the details of the numerical simulations that demonstrate this process.

\section{Model}\label{Model}

Modelling the full quantum field of a BEC in the presence of interactions between the particles leads to a model that is extremely difficult to solve either analytically or numerically.  Fortunately, the pathological case in the semiclassical model was not the interacting case, but the linear case \cite{haine04}.  In the absence of interactions there were non-stationary states that were not cooled by the feedback.  By contrast, the presence of interactions coupled these states to others that were affected by the feedback, and the system approached the ground state.  We investigate the linear case therefore, where the BEC is equivalent to a single trapped particle, and the full quantum system is tractable.

The trapping potential for the BEC is modelled as a harmonic potential of angular frequency $\omega$, with a controllable linear potential $u(t)q$.  In the absence of interactions, the Hamiltonian is given by: 
\begin{eqnarray}
H_A & = & \frac{p^2}{2m} + \frac{1}{2} m\omega^2 q^2 - u(t)q ,
\label{HA}
\end{eqnarray} 
where $u(t)$ is the control signal, to be determined.  As the Hamiltonian is quadratic in system operators and bilinear in system operators and the control signal, it is expressed in the more compact matrix form 
\begin{eqnarray}
H_A = \frac{1}{2} \xop^T \mathbf{G} \xop - \xop^T\mathbf{\Sigma} \mathbf{B} \mathbf{u},
 \label{HAx}
\end{eqnarray} 
where $\mathbf{G}$ is real and symmetric, $\mathbf{B}$ is real and $\xop$ is a vector of system operators such that the commutation relations are represented by $\xop\xop^T - (\xop\xop^T)^T \ = \ i\hbar\mathbf{\Sigma}$.  For the system considered here these become
\begin{eqnarray}
\xop &=&  \left(\begin{array}{c} q \\ p \end{array}\right), \: 
\mathbf{\Sigma} =  \left(\begin{array}{cc} 0 & 1 \\ -1 & 0 \end{array}\right), \nonumber \\ 
\mathbf{Bu} &=& \left(\begin{array}{c}0\\u(t)\end{array}\right), \:\mathbf{G}  =  \left(\begin{array}{cc} m\omega^2&0\\0&1/m \end{array}\right).
\label{vecx}
\end{eqnarray}

The feedback control of this system via a process of continuous measurement requires a model of the interaction with the measurement device. There are many treatments in the literature of how to introduce a measurement on a quantum system in different contexts~\cite{davies_76,barchielli_86,gisin_84,diosi_86,caves_87,belavkin_87,wise_milb_homo93,carm_bk}. We consider the interaction between the system and the probe field to be represented by the coupling operator $\hat L=\mathbf{L}^T\xop$, such that the master equation for the continuously monitored system will then be
\begin{eqnarray}
d\rho &=& \frac{-i}{\hbar}\left[H_A,\rho\right]\;dt +\mathcal{D}[\hat L]\rho\;dt \nonumber \\
&=& \frac{-i}{\hbar}\left[H_A,\rho\right]\;dt + \a \mathcal{D}[q]\rho\;dt \, , 
\end{eqnarray}
where $\mathcal{D}[c]\rho=c \rho c^\dagger -1/2 (c^\dagger c \rho + \rho c^\dagger c)$ and the definition
\begin{eqnarray*}
\mathbf{L} & = & \left(\begin{array}{c} \sqrt{\alpha} \\ 0 \end{array}\right), 
\end{eqnarray*}
for $\mathbf{L}$, where $\a$ is a real-valued number, has been used in the second line. This represents a system undergoing a continuous position measurement, which can be obtained, for example, by placing a BEC in an optical cavity  \cite{doherty99}. Here $\alpha$ is the measurement strength and dependent on physical parameters of the probing system (e.g., optical cavity) but is left unspecified. It is also formally possible to extract position measurements from system-reservoir couplings that equate to pure damping of the motional state, but it is precisely the absence of such cooling mechanisms that motivates our search for control schemes.  This position measurement produces momentum diffusion in the system, and we will show that this momentum diffusion makes the ground state an attractor for the system under feedback control.  

Measurement of the real quadrature of the outgoing probe field provides a measurement signal $dq_{meas}$ upon which the conditional state $\pi_t$ is updated.  Thus the conditional density operator $\pi_t$  is a function of the measurement signal, and provides a best estimate $\Xt$ of any system operator $X$ conditioned on those measurement results (in the mean square sense):  
\begin{eqnarray*}
\Xt & = & \Tr[\pi_t X].
\end{eqnarray*}
The ensemble average $\mathbf{E}[\Xt]$ is equal to the expectation calculated from the entire model $\langle X_t\rangle$:
\begin{eqnarray}
\mathbf{E}[\Xt] \ = \ \Tr[\rho_0X_t] \ \equiv \ \langle X_t\rangle,
\label{Epi_E}
\end{eqnarray}  
so that $\mathbf{E}[\pi_t]=\rho_t$ i.e. the classical average of the conditional density operator $\pi_t$ is equal to the unconditional density operator $\rho_t$.
Using either the reference probability technique~\cite{belav_92,bouten_05}, or the innovations method~\cite{belav_92a,bouten_06}  (see also~\cite{vanhandel_05,vanhandel_05b}),  the stochastic master equation for the quantum filter is
\begin{eqnarray}
d\pi_t =&&\frac{-i}{\hbar}\left[H_A,\pi_t\right]\;dt + \a \mathcal{D}[q] \pi_t dt \nonumber \\ &&+ \sqrt{\eta\a} \left(q\pi_t + \pi_tq - 2\pi_t\Tr[q\pi_t ]\right)\;dW\label{dpi} ,
\end{eqnarray}
where $dW/2\sqrt{\eta \alpha} = dq_{meas} - \qt dt \ = \ dq_{meas} - \Tr[\pi_tq]dt$, $dW$ is gaussian white noise (a Wiener process) and $\eta$ is the detection efficiency. The stochastic term in the above equation represents the additional noise introduced by the measurement signal $dq_{meas}$ due to the finite strength of the continuous measurement. 

A control signal $u(t)$ is fed back into the position of the minimum of the trap potential, and this involves a term in the Hamiltonian of the form
$
H_{control} \ = \ u(t) q. \label{control}
$
Note that for arbitrary system operator $X$ the dynamical equation for the conditional expectation $\breve{X}_t$ has a term due to the control 
\begin{eqnarray}
\mbox{Tr}\left[\frac{i}{\hbar} \left[H_{control},\pi_t\right]\;X \right]  
& = & \frac{i}{\hbar}\;u(t)\,\Tr[\,[X,q]\pi_t]  .
\label{control_arguement}
\end{eqnarray}
Thus a control proportional to $q$ affects an operator $X$ only if the two do not commute, and controlling the trap position via the term $u(t)\; q$ involves feedback to the momentum quadrature. 

We now define the control law $u(t)$ as a simple feedback of the conditional expectation of position and moment $\qt, \pt$ in a linear form:
\begin{eqnarray*}
u(t) \ = \ k_q \qt + k_p\pt.
\end{eqnarray*}  
This is s state-estimate feedback, in line with the choice used by \cite{haine04} who feed back the position moment, and is similar to that chosen by \cite{doherty99} except that the requirement of optimality has been relaxed.  Using the vector notation of Eq.(\ref{vecx}) it is fruitful to define a vector of conditional means 
\begin{eqnarray}
\xt & = & \Tr[\xop \; \pi_t] ,
 \label{xt}
\end{eqnarray}
so that the control term in the Hamiltonian can be expressed as
\begin{eqnarray}
\mathbf{Bu} = \left(\begin{array}{c}0\\u(t)\end{array}\right)
= \left(\begin{array}{cc} 0 & 0 \\ k_q & k_p \end{array}\right)\xt
 \equiv  \mathbf{K}\xt.  
\label{Bu}
\end{eqnarray}

\section{Final expected energy}\label{FEE}

The objective for this control scheme is to drive the energy of the system towards the ground state, and the effectiveness of the control can be discerned from the final energy of the system.  Since this is a quantum stochastic formulation, the expectation, or averaging, over all possible quantum trajectories corresponding to all possible measurement outcomes is required.  Moreover for a general state (other than an energy eigenstate) the final energy will be a random variable that will require an expectation.  This quantity can calculated from the energy operator in the Schrodinger picture given by
$
E  =  \frac{1}{2}\xop^T \mathbf{G} \xop 
$  
and the quantum filter $\pi_t$, using the property
$
\langle E_t \rangle  
=  \mathbf{E}[\,\Et]
=  \mathbf{E}\left[\Tr[\pi_t,E]\right]
$.   
The long term stochastic limit for the expected energy is of interest, which requires the stochastic master equation for the filter Eq.(\ref{dpi}) and the equation for the feedback term given by Eq.(\ref{Bu}).   This limit for a Gaussian state is analytically calculated below, and an argument that all initial states will be driven towards this same limit is provided, along with supporting numerical simulations.
 
Insight is gained by looking at the evolution of a Gaussian density operator under Eq.(\ref{dpi}).  If an initial state is Gaussian then it remains Gaussian,  due to the linearity of the coupling, the quadratic form of the Hamiltonian and the quantum Gaussian noise. 

Assuming a Gaussian form for the density operator $\pi_t$, it can be described by a vector of means $\xt$ given by Eq.(\ref{xt}) and a matrix of symmetrized covariances $\mathbf{V}_t$ given as 
\begin{eqnarray}
\mathbf{V}_t 
& = & \frac{1}{2}\mbox{Tr}\left[\left(\xop\xop^T + [\xop\xop^T ]^T\right)\pi_t\right] - \xt \xt^T. \label{Y_defn}
\end{eqnarray} 
Using Eq.(\ref{dpi}) and the control law Eq.(\ref{Bu}) the dynamical equations for the means and symmetrized covariances are calculated in their matrix form and come out as 
\begin{eqnarray}
d\xt & = & 
(\mathbf{A+K})\xt\;dt + \sqrt{\eta}\;\left(2\mathbf{V}_t\mathbf{L}\right)\;dW\label{meansK}\\
\dot{\mathbf{V}}_t & = & \mathbf{AV}_t + \mathbf{V}_t\mathbf{A}^T + \mathbf D - 4\eta \mathbf{V}_t \mathbf{L}(\mathbf{L})^T\mathbf{V}_t ,
 \label{vars}
\end{eqnarray}
where $\mathbf{D} \ = \ \hbar^2\mathbf{\Sigma}\left[\mathbf{L}(\mathbf{L})^T\right]\mathbf{\Sigma}^T$ and $\mathbf{A} \ = \ \mathbf{\Sigma G}$, and Eq.\ref{vars} is often referred to as a Riccati equation.  Using the definition of the covariance matrix $\mathbf{V}_t$ Eq.(\ref{Y_defn}) the average of the conditional energy provides the expected energy of 
\begin{eqnarray}
\langle E_t \rangle 
& = & \frac{1}{2} \mathbf{E}\left[ \xt^T\mathbf{G}\xt  \right]  + \frac{1}{2} \Tr[\mathbf{G}\mathbf{V}_t]  .
\label{E_t}
\end{eqnarray}
The expected energy is proportional to the expectation of the square of both position and momentum operators, which, for a Gaussian operator, is given by the variance plus the means squared.  

It is possible to analytically solve Eq.(\ref{E_t}) in the limit $t \rightarrow \infty$.  Eq.(\ref{meansK}) for the means is a multivariate Ornstein Uhlenbeck process with time dependent parameters \cite{gardiner_bk}.  Note that the variance matrix $\mathbf{V}_t$ changes with time.  For a negative value of $k_p$ the eigenvalues of $(\mathbf{A+K})$ has strictly negative real parts, meaning it is a stable or Hurwitz matrix.  This implies the dynamical system Eq.(\ref{meansK}) is stable and $\exp[(\mathbf{A+K})t] \rightarrow 0$ as $t \rightarrow  \infty$.  The solution is formally written as
\begin{eqnarray}
\xt =&&\exp\left[(\mathbf{A+K})t\right]\xo \nonumber \\ &&+ 2\sqrt{\eta}\int_0^t\exp\left[(\mathbf{A+K})(t-s)\right]\mathbf{V}_s\mathbf{L}dW_{s}. 
\label{means_solution}
\end{eqnarray}
Since having $k_q \neq 0$ does not provide any damping of the means, for the rest of the paper it is assumed $k_q = 0$ and $k_p< 0$.  This is easily interpreted: a control signal of a negative momentum estimate, coupled to the position operator, feeds back into the momentum estimate, and thus the control provides damping of the momentum, and both of the conditional means. 

Convergence of the Riccati equation Eq.(\ref{vars}) to a unique non-negative definite stabilizing solution is assured in our case\cite{davis_bk}, and can be obtained by solving the algebraic Riccati equation, $\dot{\mathbf{V}}_\infty = 0$:  
$$
\mathbf{AV}_\infty + \mathbf{V}_\infty \mathbf{A}^T + \mathbf D - 4\eta \mathbf{V}_\infty \mathbf{L}(\mathbf{L})^T\mathbf{V}_\infty  = 0.
$$
It is fortunate that this equation has an analytical solution \cite{doherty99}, 
$$
\mathbf{V}_\infty = \left(
\begin{array}{cc}
V^{qq}_\infty & V^{qp}_\infty
\\
V^{qp}_\infty & V^{pp}_\infty
\end{array}
\right)  ,
 $$
where the steady state for variances are given as \cite{doherty99}
\begin{eqnarray*}
V^{qq}_\infty &=&  \left(\frac{\omega}{2\sqrt{2}\alpha\eta}\right)\sqrt{\xi - 1} \\
V_{\infty}^{pp} 
& = & \left(\frac{m^2\omega^3}{2\sqrt{2}\alpha\eta}\right)\xi \sqrt{\xi - 1}\\
V^{qp}_{\infty} 
& = & \left(\frac{m\omega^2}{4\alpha\eta}\right)(\xi-1) \\
\xi 
& = & \sqrt{1 + \frac{4\hbar^2\eta\ \alpha^2}{m^2\omega^4}}. 
\end{eqnarray*}
It is easy to check these variances obey the Heisenberg Uncertainty principle, and the purity of the state can be calculated as \cite{doherty99, Zureck} 
\begin{eqnarray*}
\Tr[\pi_\infty^2] & = & \frac{\hbar}{2}\left(V^{qq}_\infty V^{pp}_\infty - (V^{qp}_\infty)^2\right)^{-1/2} \ = \ \sqrt{\eta}.
\end{eqnarray*}

Expanding out the variances to 2nd order for small $\alpha$ become
\begin{eqnarray*}
V^{qq}_\infty(\alpha) & \approx & \frac{\hbar}{2m\omega}\left(1-\frac{\eta\hbar^2}{2m^2\omega^4}\;\alpha^2\right)\ , \\
V^{pp}_\infty(\alpha) & \approx & 
\frac{\hbar m\omega}{2\sqrt{\eta}} \left(1+\frac{3\eta\hbar^2}{2m^2\omega^4}\;\alpha^2\right) \ , \\
V^{qp}_\infty(\alpha)& \approx & \frac{\hbar^2}{2m\omega^2}\;\alpha\  .
\end{eqnarray*}

Thus the steady state limit $t \rightarrow \infty$ of the means and variances are available.  Since the Gaussian state can be described by these parameters and their limits, the stochastic steady state limit under Eq.(\ref{dpi}) for a Gaussian state is possible.  However, using just the long term limits of both the variances Eq.(\ref{vars}) and the expectation of the square of the conditional means Eq.(\ref{means_solution}) \cite{gardiner_bk} the limit for expected energy Eq.(\ref{E_t}) comes out analytically as
\begin{eqnarray}
\label{Ess} 
\langle E_\infty \rangle 
 = &&\frac{1}{2} m\omega^2 V_{\infty}^{qq} + V_{\infty}^{pp}/2m \nonumber \\ &&+ \eta\alpha\left( 2V_{\infty}^{qq}V_{\infty}^{qp} -mk_p(V_{\infty}^{qq})^2 - \frac{\hbar^2}{k_p2\eta m}\right).
\end{eqnarray}
 
\begin{figure}[h!t]
\begin{center}
  \includegraphics[width=3in,height=2in]{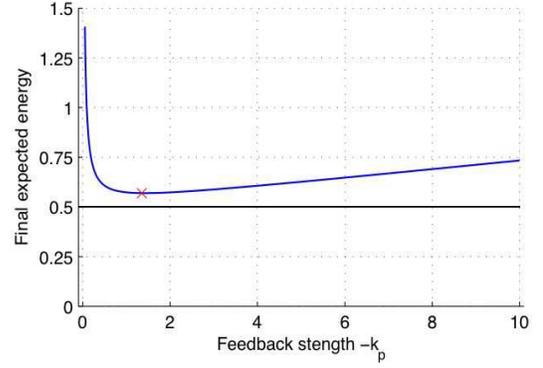}  
  \caption{Final expected energy, Eq.(\ref{Ess}), as a function of the feedback parameter $k_p$, for $\hbar = m = \omega = \eta = 1$ and $\alpha = 0.09$.  The horizontal line represents the ground state energy, while the cross represents the calculated optimal control strength.}\label{pic_feedback}
\end{center}
\end{figure}

As $\langle E_\infty \rangle$ has terms that are both linear and inversely proportional to the feedback strength $k_p$, there exists an optimal negative value for the strength of the feedback, as shown in Fig.\ref{pic_feedback} and given by
\begin{eqnarray*}
k_p^{opt} & = & -\frac{\hbar^2}{2\eta m^2}(V^{qq}_\infty)^{-2}\,.
\end{eqnarray*}

\begin{figure}[h!t]
\begin{center}
  \includegraphics[width=3in,height=2in]{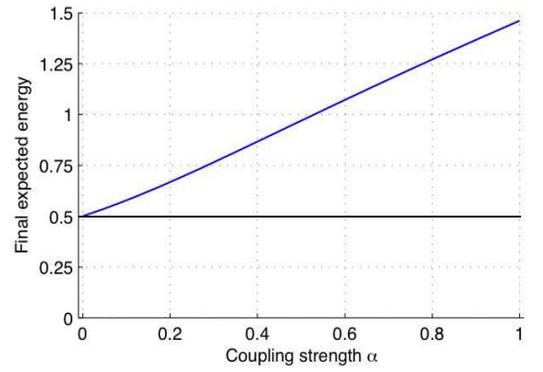}
  \caption{Final expected energy, Eq.(\ref{Ess}), as a function of coupling strength $\alpha$, for $k_p = -1.35$. The horizontal line represents the ground state energy.}\label{pic_coupling}
\end{center}
\end{figure}

For all the numerical simulations shown in this paper we used $\hbar=m=\omega=\eta=1$ and therefore all quantities displayed in the figures have dimensions compatible with this choice. 
Fig.\ref{pic_coupling} shows the relationship between the final expected energy and the coupling strength between the probe and system.  It highlights the fact that a weaker coupling introduces a smaller disturbance to the system, which can also be seen for the  expansion to 2nd order for small $\alpha$ of $\langle E_\infty \rangle$ 
\begin{eqnarray}
\langle E_\infty \rangle (\alpha) & \approx & \frac{\hbar\omega}{2}\left(1 + \frac{1}{2\sqrt{\eta}}\right) 
 - \frac{\hbar^2}{4m\omega^2}\left(k_p\eta + \frac{2\omega^2}{k_p}\right)\alpha \nonumber\\&& 
 +  \frac{\hbar^3\sqrt{\eta}}{4m^2\omega^3}\left(\sqrt{\eta} + 3/2\right)\;\alpha^2 .
\end{eqnarray}  


If the initial state $\pi_0$ is not Gaussian, then Eq.(\ref{Ess}) can still be expected to hold. While a detailed proof of this is beyond the scope of the present paper, evidence supporting this is given in the simulations in the following section. In fact, the limit conditional state is expected to be Gaussian, with zero mean and covariance $\mathbf{V}_{\infty}$. Such limits were investigated in \cite{Chrus_Stas}, and in the classical case in \cite{ocone_96}. 

To see why this might be true, we expand an arbitrary initial state $\pi_0$ as a sum of Gaussian states, i.e. a $P$ representation over the coherent states $|\beta\rangle$:
\begin{eqnarray}
\pi_0 & = &\int  P(\beta)\;|\beta\rangle\langle\beta| d\beta .
\label{sigma_expansion}
\end{eqnarray}

The stochastic master equation for a state Eq.(\ref{dpi}) is not linear in $\pi_t$, so it is not possible to apply the stochastic master equation to each state individually.  Fortunately there exists an equation for the evolution of the unnormalized state $\sigma_t$ that is linear in $\sigma$.  This is the equation for the unnormalized Belavkin filter  \cite{belav_92}, the quantum counterpart for the Zakai equation from classical filtering,
\begin{eqnarray}
d\sigma_t =&& 
\frac{-i}{\hbar}\left[H_A,\sigma_t\right]\,dt + \alpha(q\sigma_tq - \frac{1}{2} q^2\sigma_t - \frac{1}{2}\sigma_tq^2)\,dt \nonumber \\ &&+ 2\sqrt{\eta}\alpha\left(q\sigma_t + \sigma_tq \right)\,dq_{meas}, 
\label{dsigma}
\end{eqnarray}
where $dq_{meas}$ is a measurement signal driving the system, and $\sigma_0=\pi_0$. In quantum filtering this equation is often derived first, and the stochastic master equation Eq.(\ref{dpi}) for the normalised state $\pi_t$ is easily achieved through a process of normalization: here $\pi_t=\sigma_t/n_t$, where $n_t = \Tr[\sigma_t]$. 

Let $\sigma_t^\beta$ be evolution of the individual coherent initial state $\vert \beta \rangle \langle \beta \vert \equiv \sigma^\beta_0$ under Eq.\ref{dsigma}, with normalisation factor $n^\beta_t=\Tr[\sigma^\beta_t]$ and corresponding normalised state $\pi^\beta_t=\sigma^\beta_t/n^\beta_t$. By linearity the unnormalised states $\sigma_t^\beta$ can be summed, and the total then normalised, to end up with 
\begin{eqnarray}
\pi_t & = & \frac{\int P(\beta)\sigma_t^\beta d\beta}{n_t} 
\ = \ \frac{\int P(\beta)\pi_t^\beta n_t^\beta d\beta}{n_t}
\label{pi_expansion}
\end{eqnarray}
where the total normalisation factor is given by $n_t = \Tr[\int P(\beta)\sigma_t^\beta d\beta ] = \int P(\beta)n_t^\beta d\beta$.

Using the classical limiting results for gaussian initial conditions \cite{ocone_96},  for large times $\pi_t^\beta$ is independent of $\beta$ i.e. $\pi_t^\beta \approx \pi_t^{\beta_\infty}$ for all $\beta$, where $\pi_t^{\beta_\infty}$ is the limiting state given by a gaussian of mean $\breve{x}_t$ and variances $\mathbf{V}_\infty$.  Thus for large times we have 
\begin{eqnarray}
\pi_t & \approx & \frac{\pi_t^{\beta_\infty}\int P(\beta) n_t^\beta d\beta}{n_t} \ = \ \pi_t^{\beta_\infty}.
\label{pi_solution}
\end{eqnarray}

\section{Simulation} \label{Sim}

When simulating the system, the stochastic master equation for the quantum filter Eq.(\ref{dpi}) is expressed as a stochastic differential equation for a $Q$ function \cite{GZ} with the relation 
$$
dQ(x,y,t) \ = \ \frac{1}{\pi}\,\langle \,x + i y \,| \,d\pi_t \,|\, x + i y \,\rangle  ,
$$
where $|x+i y\rangle$ is a coherent state, so that
\begin{widetext}
\begin{eqnarray}
dQ(x,y,t) & = & 
\left(\omega\left(x\frac{\partial}{\partial y} - y\frac{\partial}{\partial x}\right) - k_p \left(\int\int dx'dy'\  y'Q(x',y',t)\right)\frac{\partial}{\partial y} + \frac{\hbar\a}{4m\omega}\frac{\partial^2}{\partial y^2}\right)Q(x,y,t)\;dt \nonumber \\&& + 2\a\sqrt{\frac{\hbar\eta}{2m\omega}}\;\left(4x + \frac{\partial}{\partial x} - 4\left(\int\int dx'dy'\; x'Q(x',y',t)\right)\right)Q(x,y,t)\;dW .
 \label{Q_xy}
\end{eqnarray}
\end{widetext}

The first term, proportional to $\omega$, causes a rotation in phase space around the origin due to the harmonic potential. The second term is the effect of the control, proportional to $k_p$ and the mean momentum estimate, and drives the $Q$ function towards the origin.  The third is the backaction of position measurement, which involves a broadening of the Gaussian in the momentum quadrature.  The last part is called the innovation term, and involves the Wiener increment $dW$ arising from the random update of the best estimate coming from the measurement data. It causes the function to be stochastically shifted around the phase space as well as a narrowing of the state in the position quadrature.

\begin{figure}
\includegraphics[width=7.5cm]{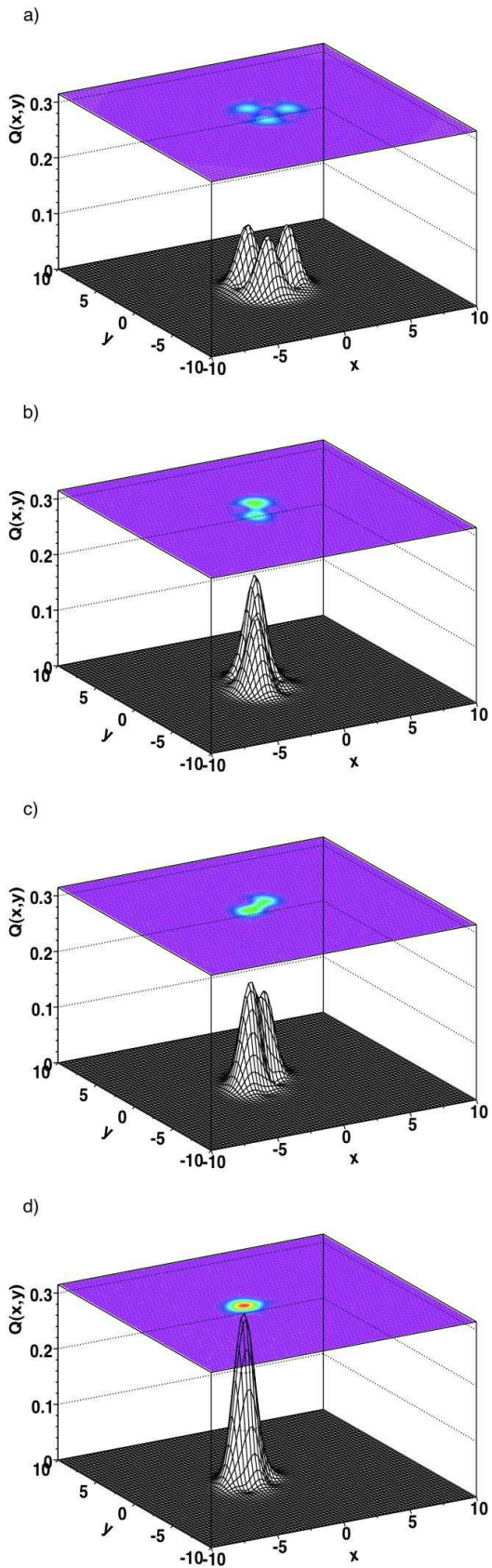}
\caption{$Q$-functions at different times (corresponding to the letters in Fig.~\ref{energytime}) for the evolution of the dark state $\pi_0$ (a) under Eq.(\ref{Q_xy}), for $k_p=-1.35$ and $\alpha=0.3$.}\label{timeevolQ}
\end{figure}

\begin{figure}
\includegraphics[width=8cm]{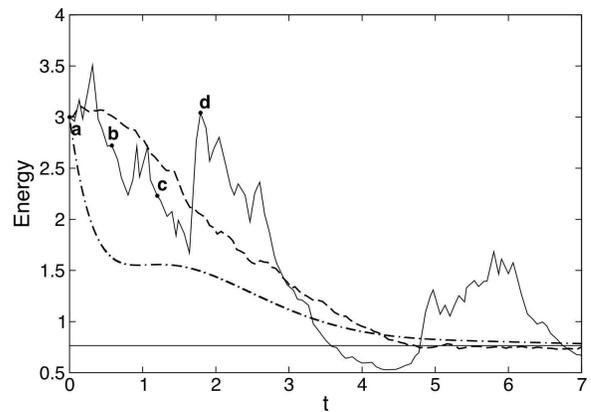}
\caption{Time evolution of the expected energy for the same parameters as in Fig.~\ref{timeevolQ}. Solid line represents a single trajectory, dashed line is the result after averaging over 48 realisations of the noise in Eq.(\ref{Q_xy}), and dot-dashed line is the solution for an initial Gaussian state (see Eq.~(\ref{E_t})). The horizontal line corresponds to the steady state solution given by Eq.~(\ref{Ess}) and the letters to the times at which the $Q$-functions in Fig.~\ref{timeevolQ} are plotted.}\label{energytime}. 
\end{figure}

Equation (\ref{Q_xy}) was simulated in a Stratonovich form using the numerical package XMDS~\cite{xmds}. The initial condition was chosen to be a linear combination of two number states, $\pi_0  = \vert \psi_0\rangle \langle \psi_0 \vert$, with $\vert \psi_0\rangle = (\vert 1 \rangle +\vert 4 \rangle )/\sqrt{2}$, which satisfies the dark state conditions, where in the absence of measurement back-action the feedback would have  had no effect on the energy of the system \cite{haine04}. 
The combined effect of all the terms in Eq.~(\ref{Q_xy}) in the dynamics can be seen in Fig.~\ref{timeevolQ} where the $Q$ function at various times is plotted for $k_p =-1.35$ and $\a = 0.3$. As argued in the previous section, as a result of the measurement process itself the initial dark state (Fig.~\ref{timeevolQ}-a) continuously turns into a Gaussian, becoming subject to the effects of the control, which will cool the system down.
This cooling dynamics is shown in Fig.~\ref{energytime} for the same parameters as in Fig.~\ref{timeevolQ}. For a single realisation of the noise in Eq.~(\ref{Q_xy}) the energy will asymptotically fluctuate around the steady state solution of Eq.~(\ref{Ess}) (horizontal line). The figure also shows the result averaged over 48 realisations (dashed line) where the energy starts to converge to its asymptotic predicted value. A curve with the solution, Eq.(\ref{E_t}), for a given Gaussian initial state is also shown for comparison (dot-dahsed line): cooling is effective since the beginning for a initial Gaussian  while it takes longer for an initial dark state. 

\section{Conclusion}\label{Conclusion}

This paper shows that a quantum system under continuous measurement, in a harmonic potential and driven by simple feedback law, will evolve towards a stable, low-energy state, and that this occurs for all initial states.  In the limit of weak measurement, the steady state is the ground state of the system.  This can be deduced from the idea that all initial states under continuous measurement evolve towards a Gaussian state \cite{Chrus_Stas}, \cite{Belavkin_Staszewski_92}. We provide an analytical result for the steady state limit for a Gaussian state under continuous measurement and subject to the control law suggested here, which we then infer to be the steady state for all initial states. This result opens a new possibility to control the spatial mode of a BEC as it eliminates concern raised by Haine \emph{et al.} \cite{haine04} about dark states: energetic states that do not display an error signals and hence are not susceptible to control.  These states are no longer stable under the measurement and control scheme suggested here, and thus can be controlled and driven towards the ground state.

The analytic results are supported by numerical simulations based on a phase-space representation of the quantum state.  These simulations are general in that they allow any quantum state for the system and a wide range of allowable evolution, but are only tractable for single mode problems. These numerical simulations can provide the basis for extensions of this linear model to non-linear systems. 

The stochastic steady state has a minimum energy at the optimal feedback strength, and the overall energy of the steady state is limited by the measurement strength. Setting the measurement strength to a infinitesimal value is counter-productive in practice, of course, as the time-scale for the convergence to the steady state becomes infinitely long.  In practice it may be useful to adjust the strength of the measurement to provide rapid convergence and then a low steady-state energy, but this will be examined in future work.

\linespread{1.0}
\bibliography{qcontrol}

\end{document}